\documentclass[12pt]{article}
\usepackage[english]{babel}
\usepackage{graphicx}

\begin{document}

\begin{center}
\begin{large}
\textbf{The Schr$\mathrm{\ddot{o}}$dinger-Foldy relativistic canonical quantum mechanics and the derivation of the Dirac equation}
\end{large}
\vskip 0.5cm

\textbf{I.Yu. Krivsky$^1$, V.M. Simulik$^1$, I.L.
Lamer$^1$, T.M. Zajac $^2$}
\end{center}

\begin{center}
\textit{$^1$Institute of Electron Physics, National Academy of
Sciences of Ukraine, 21 Universitetska Str., 88000 Uzhgorod,
Ukraine; E-mail: vsimulik@gmail.com}

\end{center}

\begin{center}
\textit{$^2$Uzhgorod National University, Department of
Electronic Systems, 13 Kapitulna Str., 88000 Uzhgorod, Ukraine}
\end{center}

\vskip 1.cm

\noindent ABSTRACT. The detailed consideration of the
relativistic canonical quantum-mechanical model of an arbitrary
spin-multiplet is given. The group-theoretical
analysis of the algebra of experimentally observables physical
quantities for the spin 1/2 doublet is presented. It is shown that both the
Foldy-Wouthuysen equation for the fermionic spin 1/2
doublet and the Dirac equation in its local representation are
the consequences of the relativistic canonical quantum mechanics
of the corresponding doublet. The mathematically
well-defined consideration on the level of modern axiomatic
approaches to the field theory is provided. The hypothesis of the antiparticle negative mass is discussed briefly in the section conclusions. 

\vskip 0.5cm


\vskip 0.5cm


\vskip 1.cm

\section{Introduction}

The extended and detailed presentation of the results of the paper
[1] is given. The basic principles of relativistic canonical quantum mechanics (RCQM) for the spin $s=\frac{1}{2}$ doublet and the derivation of the Dirac equation from this model are under
further consideration. The foundations of RCQM were given in [2]-
[4] and a procedure of axiomatic construction of this theory was
shown briefly in [1]. Here the mathematically
well-defined consideration on the level of modern axiomatic
approaches to the field theory [5] is provided.

The significance of the Dirac equation and its wide-range
application in different models of theoretical physics (QED, QHD,
theoretical atomic and nuclear physics, solid systems) is
well-known. The recent application of the massless Dirac equation
to the graffen ribbons is an example of possibilities of this
equation. Therefore, the new ways of the derivation of the Dirac
equation are the interesting problems.

Here we consider a problem whether there exists a model of a
"particle doublet" (as an elementary fundamental object), from which
the Dirac equation would follow directly and unambiguously. We
are able to demonstrate that axiomatically formulated RCQM of a
particle-antiparticle doublet of spin $s=\frac{1}{2}$ should be
chosen as such a model. The illustration of this assertion on the
example of electron-positron doublet, $e^{-}e^{+}$-doublet, is
given.

The model of RCQM for the elementary particle with $m>0$ and spin $s=\frac{1}{2}$, which satisfy the equation $i\partial_{t}\varphi(x)=\sqrt{m^{2}-\Delta}\varphi(x); \, x\in \mathrm{M}(1,3), \, \int d^{3}x\left|\varphi(x)\right|^{2}<\infty$, was suggested and approved in [2] - [4]. This model can be easily generalized to the case of arbitrary $\overrightarrow{s}$-multiplet, i. e. the "elementary object" with mass $m$ and spin $\overrightarrow{s} \equiv\left(s^{j}\right)=\left(s_{23},s_{31},s_{12}\right): \, \left[s^{j},s^{l}\right]=i\varepsilon^{jln}s^{n}$, where $\varepsilon^{jln}$ is the Levi-Civita tensor and $s^{j}=\varepsilon^{j\ell n}s_{\ell n}$ are the Hermitian $\mathrm{M}\times \mathrm{M}$ matrices -- the generators of M-dimensional representation of the spin group SU(2) (universal covering of the SO(3)$\subset$SO(1,3) group).

In this article we present the detalization of such generalization at the example of the spin $s=\frac{1}{2}$ Fermionic doublet. All mathematical and physical details of consideration (e. g. the algebras of all experimentally observables physical quantities, related to the choice of the concrete form of the spin $\overrightarrow{s}$ doublet, at the example of $e^{-}e^{+}$-doublet are illustrated.

At first we have presented the main conceptions of RCQM. Further, the group-theoretical
analysis of the algebra of observables is fulfilled. The detailed
consideration of the basic set of operators, which completely
determine the algebra of all experimentally observables physical
quantities, at the example of $e^{-}e^{+}$-doublet is given. The
special role of the stationary complete sets of corresponding
operators of observables is demonstrated. Finally, we have found the
operator, which translates the equation and the algebra of
observables of RCQM into the equation and the algebra of
observables of the Foldy-Wouthuysen (FW) representation for the
spinor field. We have found also the operator, with the help of
which the Dirac equation in its local representation and the
corresponding algebra of observables are derived directly from the
equation of motion of RCQM and from the algebra of observables in
this model.

We choose here the standard relativistic concepts, definitions and notations
in the convenient for our consideration form. For example, in the
Minkowski space-time

\begin{equation}
\label{eq1}
\mathrm{M}(1,3)=\{x\equiv(x^{\mu})=(x^{0}=t, \,
\overrightarrow{x}\equiv(x^{j}))\}; \quad \mu=\overline{0,3}, \,
j=1,2,3,
\end{equation}

\noindent the $x^{\mu}$ are the Cartesian (covariant)
coordinates of the points of the physical space-time in the
arbitrary-fixed inertial frame of references (IFR). We use the system of units $\hbar=c=1$. The metric tensor is given by

\begin{equation}
\label{eq2}
g^{\mu\nu}=g_{\mu\nu}=g^{\mu}_{\nu}, \, \left(g^{\mu}_{\nu}\right)=\mathrm{diag}\left(1,-1,-1,-1\right); \quad x_{\mu}=g_{\mu\nu}x^{\mu},
\end{equation}

\noindent the summation over the twice repeated index is implied.

The analysis of the relativistic invariance of an arbitrary physical model demands as a first step the consideration of its invariance with respect to the proper ortochronous Lorentz $\mbox{L}_ + ^ \uparrow $ = SO(1,3)=$\left\{\Lambda=\left(\Lambda^{\mu}_{\nu}\right)\right\}$ and  Poincar$\mathrm{\acute{e}}$ $\mbox{P}_ + ^
\uparrow = \mbox{T(4)}\times )\mbox{L}_ + ^ \uparrow  \supset \mbox{L}_ + ^ \uparrow$ groups. This invariance in an arbitrary relativistic model is the realization of the Einstein's relativity principle in the form of special relativity. Note that the mathematical correctness demands to consider the invariance mentioned above as the invariance with respect to the universal coverings $\mathcal{L}$ = SL(2,C) and $\mathcal{P}\supset\mathcal{L}$ of the groups $\mbox{L}_ + ^ \uparrow $ and $\mbox{P}_ + ^ \uparrow $, respectively.

For the group $\mathcal{P}$ we choose the real parameters $a=\left(a^{\mu}\right)\in$M(1,3) and $\varpi\equiv\left(\varpi^{\mu\nu}=-\varpi^{\nu\mu}\right)$, which physical meaning is well-known. For the standard $\mathcal{P}$ generators $\left(p_{\mu},j_{\mu\nu}\right)$ we use the commutation relations in the manifestly covariant form 

\begin{equation}
\label{eq3}
\left[p_{\mu},p_{\nu}\right]=0, \, \left[p_{\mu},j_{\rho\sigma}\right]=ig_{\mu\rho}p_{\sigma}-ig_{\mu\sigma}p_{\rho}, \, \left[j_{\mu\nu},j_{\rho\sigma}\right]=-i\left(g_{\mu\rho}j_{\nu\sigma}+g_{\rho\nu}j_{\sigma\mu}+g_{\nu\sigma}j_{\mu\rho}+g_{\sigma\mu}j_{\rho\nu}\right).
\end{equation}

\section{Group-theoretical analysis of the algebra of observables in relativistic canonical quantum mechanics of the Fermi-doublet}

The relativistic quantum mechanics in canonical form (RCQM) for the particle of $m>0$ and spin $s=\frac{1}{2}$ was suggested in [2]-[4]. The analysis of the principles of heredity and correspondence with the non-relativistic Schr$\mathrm{\ddot{o}}$dinger quantum mechanics was given. Such RCQM can be obviously generalized on the case of a multiplet with an arbitrary mass $m$ and SU(2)-spin 

\begin{equation}
\label{eq4}
\overrightarrow{s} \equiv\left(s^{j}\right)=\left(s_{23},s_{31},s_{12}\right): \, \left[s^{j},s^{l}\right]=i\varepsilon^{jln}s^{n}; \, \varepsilon^{123}=+1,
\end{equation}

\noindent where, as it was already mentioned above, $\varepsilon^{jln}$ is the Levi-Civita tensor and $s^{j}=\varepsilon^{j\ell n}s_{\ell n}$ are the Hermitian $\mathrm{M}\times \mathrm{M}$ matrices -- the generators of M-dimensional representation of the spin group SU(2) (universal covering of the SO(3)$\subset$SO(1,3) group).

Below we illustrate the generalization of the RCQM for an arbitrary mass $m$ and SU(2) spin on the test example of the electron-positron doublet as an "elementary physical system". Note that the case of arbitrary spin differs from the particular case $s=\frac{1}{2}$ only by the clarification of the SU(2) spin $\overrightarrow{s}$ operator (4). We pay the adequate attention to the mathematical correctness of the consideration. Moreover, the adequate attention is paid to the physical sense of the operators of the experimentally observed physical quantities. 

\textbf{The quantum-mechanical space of states}. The quantum-mechanical space of complex-valued 4-component square-integrable functions of $x\in\mathrm{R}^{3}\subset \mathrm{M}(1,3)$ is chosen for the Hilbert space $\mathrm{H}^{3,4}$ of the states of the doublet:  

\begin{equation}
\label{eq5}
\mathrm{H}^{3,4}=\mathrm{L}_{2}(\mathrm{R}^3)\otimes\mathrm{C}^{\otimes4}=\{f=(f^{\alpha}):\mathrm{R}^{3}\rightarrow\mathrm{C}^{\otimes4};
 \, \int
d^{3}x|f(t,\overrightarrow{x})|^{2} <\infty\} \}.
\end{equation}

\noindent where $d^{3}x$ is the Lebesgue measure in the space $\mathrm{R}^{3}\subset \mathrm{M}(1,3)$ of the eigenvalues of the position operator $\overrightarrow{x}$ of the Cartesian coordinate of the doublet in an arbitrary-fixed inertial frame of reference (IFR). In (5) and below the two upper components $f^{1}, \, f^{2}$ of the vector $f\in\mathrm{H}^{3,4}$ are the components of the electron wave function $\varphi_{-}$ and the two lower components $f^{3}, \, f^{4}$ are those of the positron wave function $\varphi_{+}$.

\textbf{The Schr$\mathrm{\ddot{o}}$dinger-Foldy equation of motion}. The equation of motion of the particle doublet in the space (5) (i. e. the dependence of vectors $f\in\mathrm{H}^{3,4}$ from the time $t=x^{0}$ as the evolution parameter) is determined by the energy operator of the free doublet

\begin{equation}
\label{eq6}
\widehat{\omega} \equiv \sqrt{\widehat{\overrightarrow{p}}^{2} + m^2} =\sqrt { - \Delta + m^2}\geq m>0; \quad \widehat{\overrightarrow{p}}\equiv(p^{j})=-i\nabla, \quad \nabla\equiv(\partial_{\ell}).
\end{equation}

In the $\overrightarrow{x}$-realization (5) of the space $\mathrm{H}^{3,4}$ the canonically conjugated coordinate $\overrightarrow{x}$ and momentum $\overrightarrow{p}$ satisfy the Heisenberg commutation relations

\begin{equation}
\label{eq7}
\left[x^{j},\widehat{p}^{l}\right]=i\delta^{jl}, \quad \left[x^{j},x^{l}\right]=\left[\widehat{p}^{j},\widehat{p}^{l}\right]=0,
\end{equation}

\noindent and commute with the spin operator $\overrightarrow{s}$ (4) (the explicit form of the operator $\overrightarrow{s}$ for the $e^{-}e^{+}$-doublet is detalized below in (24)). In the integral form this evolution is determined by the unitary in the space (5) operator 

\begin{equation}
\label{eq8}
u(t_{0},t)=\mathrm{exp}\left[-i(t-t_{0})\widehat{\omega}\right]; \quad \mathrm{exp}\widehat{A}\equiv\sum^{\infty}_{n=0}\frac{\widehat{A}^{n}}{n!}; \, t,t_{0}\in(-\infty,\infty),
\end{equation}

\noindent which is the automorphism operator in the space (5)
(below we put $t_{0}=0$).

In the differential form the evolution equation is given as

\begin{equation}
\label{eq9}
i\partial_{t}f(t,\overrightarrow{x})=\sqrt { - \Delta + m^2}f(t,\overrightarrow{x}), \quad f\in \mathrm{H}^{3,4}; \quad \partial_{t}\equiv\frac{\partial}{\partial t}.
\end{equation} 

\noindent Equation (9) is the equation of motion of a "particle" (doublet) in the RCQM, i. e., the main equation of the model. Moreover, we will prove below that the Foldy-Wouthuysen [2] and well-known Dirac equations are the consequences of this equation. Therefore, the equation (9) plays an outstanding role. In the papers [3], [4] the two component version of the equation (9) is called the Schr$\mathrm{\ddot{o}}$dinger equation. Taking into account the L. Foldy's contribution in the construction of RCQM and his proof of the principle of correspondence between RCQM and non-relativistic quantum mechanics, we propose \textit{to call the $N$-component equations of the type (9) as the Schr$\mathrm{\ddot{o}}$dinger-Foldy (SF) equations}. 

The pseudo-differential (non-local) operator (6) is determined alternatively either in the form of the power series

\begin{equation}
\label{eq10}
\widehat{\omega}=m\sqrt{1-\widehat{B}}\equiv 1-\frac{1}{2}\widehat{B}+\frac{1\cdot 2}{2\cdot 3}\widehat{B}^{2}-..., \quad \widehat{B}=\frac{\Delta}{m^{2}},
\end{equation}

\noindent or in the integral form

\begin{equation}
\label{eq11}
(\widehat{\omega}f)(t,\overrightarrow{x})=\frac{1}{(2\pi)^{\frac{3}{2}}}\int d^{3}k e^{i\overrightarrow{k}\overrightarrow{x}}\omega \widetilde{f}(t,\overrightarrow{k}); \quad \omega\equiv \sqrt{\overrightarrow{k}^{2}+m^{2}}, \quad \widetilde{f}\in \widetilde{\mathrm{H}}^{3,4},
\end{equation}

\noindent where $f$ and $\widetilde{f}$ are linked by the 3-dimensional Fourier transformations

\begin{equation}
\label{eq12}
f(t,\overrightarrow{x})=\frac{1}{(2\pi)^{\frac{3}{2}}}\int d^{3}k e^{i\overrightarrow{k}\overrightarrow{x}}\widetilde{f}(t,\overrightarrow{k})\Leftrightarrow \widetilde{f}(t,\overrightarrow{k})=\frac{1}{(2\pi)^{\frac{3}{2}}}\int d^{3}k e^{-i\overrightarrow{k}\overrightarrow{x}}\widetilde{f}(t,\overrightarrow{x}),
\end{equation}

\noindent (in (12) $\overrightarrow{k}$ belongs to the spectrum $\mathrm{R}^{3}_{\vec{k}}$ of the operator $\widehat{\overrightarrow{p}}$, and the parameter $t\in (-\infty,\infty)\subset\mathrm{M}(1,3)$).

Note that the space of states (5) is invariant with respect to the Fourier transformation (12). Therefore, both $\overrightarrow{x}$-realization (5) and $\overrightarrow{k}$-realization $\widetilde{\mathrm{H}}^{3,4}$ for the doublet states space are suitable for the purposes of our consideration. In the $\overrightarrow{k}$-realization the SF equation has the algebraic-differential form

\begin{equation}
\label{eq13}
i\partial_{t}\widetilde{f}(t,\overrightarrow{k})=\sqrt { \overrightarrow{k}^{2} + m^2}\widetilde{f}(t,\overrightarrow{k}); \quad  \overrightarrow{k}\in\mathrm{R}^{3}_{\vec{k}}, \quad \widetilde{f}\in \widetilde{\mathrm{H}}^{3,4}.
\end{equation}

\noindent Below in the places, where misunderstanding is impossible, the symbol "tilde" is omitted.

\textbf{On the Poincar$\mathrm{\acute{e}}$ group representation}. The generators of the $\mathcal{P}^{\mathrm{f}}$ representation of the group $\mathcal{P}$, with respect to which the equation (9) is invariant, are given by

\begin{equation}
\label{eq14}
\widehat{p}_{0}=\widehat{\omega}, \, \widehat{p}_{l}=i\partial_{l}, \, \widehat{j}_{ln}=x_{l}\widehat{p}_{n}-x_{n}\widehat{p}_{l}+s_{ln}\equiv \widehat{m}_{ln}+s_{ln},
\end{equation}

\begin{equation}
\label{eq15}
\widehat{j}_{0l}=-\widehat{j}_{l0}=t\widehat{p}_{l}-\frac{1}{2}\left\{x_{l},\widehat{\omega}\right\}-\frac{s_{ln}\widehat{p}_{n}}{\widehat{\omega}+m},
\end{equation}

\noindent in the $\overrightarrow{x}$-realization of the space $\mathrm{H}^{3,4}$ (5) and

\begin{equation}
\label{eq16}
p_{0}=\omega, \, p_{l}=k_{l}, \, \widetilde{j}_{ln}=\widetilde{x}_{l}k_{n}-\widetilde{x}_{n}k_{l}+s_{ln}; \quad (\widetilde{x}_{l}=-i\widetilde{\partial}_{l}, \, \widetilde{\partial}_{l}\equiv \frac{\partial}{\partial k^{l}}),  
\end{equation}

\begin{equation}
\label{eq17}
\widetilde{j}_{0l}=-\widetilde{j}_{l0}=t k_{l}-\frac{1}{2}\left\{\widetilde{x}_{l},\omega \right\}-\frac{s_{ln}k_{n}}{\omega+m},
\end{equation}

\noindent in the $\overrightarrow{k}$-realization $\widetilde{\mathrm{H}}^{3,4}$ of the doublet states space, respectively.

Note that the explicit form of the spin operators $s_{ln}$ in the formulae (14)-(17), which is used for the $e^{-}e^{+}$-doublet, is given in the formula (24) below.
	
In despite of manifestly non-covariant forms (14) -- (17) of the $\mathcal{P}^{\mathrm{f}}$-generators, they satisfy the commutation relations of the $\mathcal{P}$ algebra in manifestly covariant form (3). 

The $\mathcal{P}^{\mathrm{f}}$-representation of the group $\mathcal{P}$ in the space $\mathrm{H}^{3,4}$ (5) is given by the converged in this space exponential series 

\begin{equation}
\label{eq18}
\mathcal{P}^{\mathrm{f}}: \, (a,\varpi)\rightarrow U(a,\varpi)=\exp (-ia^{0}\widehat{\omega}-i\overrightarrow{a}\widehat{\overrightarrow{p}}-\frac{i}{2}\varpi^{\mu\nu}\widehat{j}_{\mu\nu}),
\end{equation}

\noindent or, in the space $\widetilde{\mathrm{H}}^{3,4}$, by corresponding exponential series given in terms of the generators (16), (17). 

We emphasize that the modern definition of $\mathcal{P}$ invariance (or $\mathcal{P}$ symmetry) of the equation of motion (9) in $\mathrm{H}^{3,4}$ is given by the following assertion, see, e. g. [6]. \textit{The set} $\mathrm{F}\equiv\left\{f\right\}$ \textit{of all possible solutions of the equation (9) is invariant with respect to the} $\mathcal{P}^{\mathrm{f}}$-\textit{representation of the group} $\mathcal{P}$, \textit{if for arbitrary solution} $f$ \textit{and arbitrarily-fixed parameters} $(a,\varpi)$ \textit{the assertion}

\begin{equation}
\label{eq19}
(a,\varpi)\rightarrow U(a,\varpi)\left\{f\right\}=\left\{f\right\}\equiv\mathrm{F}
\end{equation}

\noindent \textit{is valid}. Furthermore, the assertion (19) is ensured by the fact that (as it is easy to verify) all the $\mathcal{P}$-generators (14), (15) commute with the operator $i\partial_{t}-\sqrt { - \Delta + m^2}$ of the equation (9). The important physical consequence of the last assertion is the fact that 10 integral dynamical variables of the doublet

\begin{equation}
\label{eq20}
(P_{\mu}, \, J_{\mu\nu}) \equiv \int d^{3}x f^{\dag}(t,\overrightarrow{x})(\widehat{p}_{\mu}, \, \widehat{j}_{\mu\nu})f(t,\overrightarrow{x})=\mathrm{Const}
\end{equation}

\noindent do not depend on time, i. e. they are the constants of motion for this doublet. Below more detailed analysis of this and other meaningful assertions is presented.

\textbf{On the external and internal degrees of freedom}. The coordinate $\overrightarrow{x}$ (as an operator in $\mathrm{H}^{3,4})$ is an analog of the discrete index of generalized coordinates $q\equiv(q_{1}, q_{2}, ...)$ in non-relativistic quantum mechanics of the finite number degrees of freedom. In other words the coordinate $\overrightarrow{x}\in \mathrm{R}^{3}\subset$M(1,3) is the continuous carrier of the external degrees of freedom of a multiplet (the terminology is taken from [7]). The coordinate operator together with the operator $\widehat{\overrightarrow{p}}$ determines the operator $m_{ln}=x_{l}\widehat{p}_{n}-x_{n}\widehat{p}_{l}$ of an orbital angular momentum, which also is connected with the external degrees of freedom. 

However, the doublet has the additional characteristics such as the spin operator $\overrightarrow{s}$ (4), which is the carrier of the internal degrees of freedom of this multiplet. The set of generators $(\widehat{p}_{\mu}, \widehat{j}_{\mu\nu})$ (14), (15) of the main dynamical variables (20) of the doublet are the functions of the following basic set of 9 functionally independent operators 

\begin{equation}
\label{eq21}
\overrightarrow{x}=(x^{j}), \, \widehat{\overrightarrow{p}}=(\widehat{p}^{j}), \, \overrightarrow{s} \equiv\left(s^{j}\right)=\left(s_{23},s_{31},s_{12}\right).
\end{equation}

Note that $\overrightarrow{s}$ commutes both with $(\overrightarrow{x},\widehat{\overrightarrow{p}})$ and with the operator $i\partial_{t}-\sqrt { - \Delta + m^2}$ of the SF equation (9). Thus, for the free doublet the external and internal degrees of freedom are independent. Therefore, 9 operators (21) in $\mathrm{H}^{3,4}$, which have the univocal physical sense, are the \textit{generating} operators not only for the 10 main $(\widehat{p}_{\mu}, \widehat{j}_{\mu\nu})$ (14), (15) but also for other operators of any experimentally observable quantities of the doublet.

\textbf{On the mathematical correctness of consideration}. Note further that SF equation (9) has generalized solutions, which do not belong to the space $\mathrm{H}^{3,4}$, see the formulae (28) below. In order to account this fact it is sufficient to apply the rigged Hilbert space   

\begin{equation}
\label{eq22}
\mathrm{S}^{3,4}\equiv \mathrm{S}(\mathrm{R}^{3})\times\mathrm{C}^{4}\subset\mathrm{H}^{3,4}\subset\mathrm{S}^{3,4*}.
\end{equation}

\noindent Here $\mathrm{S}(\mathrm{R}^{3})$ is the Schwartz test function space over the space $\mathrm{R}^{3}\subset \mathrm{M}(1,3)$, and $\mathrm{S}^{3,4*}$ is the space of 4-component Schwartz generalized functions, which is conjugated to the Schwartz test function space $\mathrm{S}^{3,4}$
by the corresponding topology (see, e. g., [8]). Strictly speaking, the mathematical correctness of consideration demands to make the calculations in the space $\mathrm{S}^{3,4*}$ of generalized functions, i. e. with the application of cumbersome functional analysis.

Nevertheless, let us take into account that the Schwartz test function space $\mathrm{S}^{3,4}$ in the 	
triple (22) is \textit{kernel}. It means that $\mathrm{S}^{3,4}$ is dense both in quantum-mechanical space $\mathrm{H}^{3,4}$ and in the space of generalized functions $\mathrm{S}^{3,4*}$ (by the corresponding topologies). Therefore, any physical state $f\in\mathrm{H}^{3,4}$ can be approximated with an arbitrary precision by the corresponding elements of the Cauchy sequence in $\mathrm{S}^{3,4}$, which converges to the given $f\in\mathrm{H}^{3,4}$. Further, taking into account the requirement to measure the arbitrary value of the model with non-absolute precision, it means that all concrete calculations can be fulfilled within the Schwartz test function space $\mathrm{S}^{3,4}$.

Furthermore, the mathematical correctness of the consideration demands to determine the domain of definitions and the range of values for any used operator and for the functions of operators. Note that if the kernel space  $\mathrm{S}^{3,4}\subset\mathrm{H}^{3,4}$ is taken as the common domain of definitions of the generating operators (21), then this space appears to be also the range of their values. Moreover, the space $\mathrm{S}^{3,4}$ appears to be the common domain of definitions and values for the set of all above mentioned functions from the 9 operators (21) (for example, for the operators $(\widehat{p}_{\mu}, \widehat{j}_{\mu\nu})$ and for different sets of commutation relations). Therefore, in order to guarantee the realization of the principle of correspondence between the results of cognition and the instruments of cognition in the given model, it is sufficient to take the algebra $\mathrm{A}_{\mathrm{S}}$ of the all sets of observables of the given model in the form of converged in $\mathrm{S}^{3,4}$ Hermitian power series of the 9 generating operators (21).

\textbf{On the quantum-mechanical representation of matrix operators}. Now the qualification of the definition of the matrix operators, which describe the electron-positron $e^{-}e^{+}$-doublet, will be given. We prefer the definition, which gives the modern experimentally verified understanding of the positron as the "mirror mapping" of the electron. Such understanding leads to the specific postulation of the explicit forms of the charge sign and spin operators.

We take into account that the definition of the electron spin in the terms of the Pauli matrices is universally recognized. Therefore, we choose the electron spin in the form   
  
\begin{equation}
\label{eq23}
\overrightarrow{s}_{-}=\frac{1}{2}\overrightarrow{\sigma}, \quad \overrightarrow{\sigma}\equiv (\sigma^{j}): \, \sigma ^1 = \left| {{\begin{array}{*{20}c}
 0 \hfill & 1 \hfill \\
 1 \hfill & 0 \hfill \\
\end{array} }} \right|,\mbox{ }\sigma ^2 = \left| {{\begin{array}{*{20}c}
 0 \hfill & { - i} \hfill \\
 i \hfill & 0 \hfill \\
\end{array} }} \right|,\mbox{ }\sigma ^3 = \left| {{\begin{array}{*{20}c}
 1 \hfill & 0 \hfill \\
 0 \hfill & { - 1} \hfill \\
\end{array} }} \right|\mbox{; }j = 1,2,3. 
\end{equation}

\noindent Thus, the above mentioned understanding of positron demands to choose the sign of the charge $g$ and spin operators of the $e^{-}e^{+}$-doublet in the form

\begin{equation}
\label{eq24} g\equiv-\gamma^0 = \left| {{\begin{array}{*{20}c}
 -\mathrm{I}_{2} \hfill & 0 \hfill \\
 0 \hfill & \mathrm{I}_{2} \hfill \\
\end{array} }} \right|, \quad \bar{\overrightarrow{s}}=
\frac{1}{2}\left| {{\begin{array}{*{20}c}
 \overrightarrow{\sigma} \hfill  0 \hfill \\
 0 -C\hfill\overrightarrow{\sigma}\hfill C \\
\end{array} }} \right|, \quad \mathrm{I}_{2}=\left|
{{\begin{array}{*{20}c}
 1 \hfill & 0 \hfill \\
 0 \hfill & { 1} \hfill \\
\end{array} }} \right|,
\end{equation}

\noindent where $C$ is the operator of complex conjugation.

Indeed, only in these definitions one obtains the following result: if in the given state $f\in\mathrm{H}^{3,4}$ the electron with the charge $-e$ is in the state with the helicity value $h_{e^{-}}=-\frac{1}{2}$ (left-helical electron), then the positron is in the state $h_{e^{+}}=+\frac{1}{2}$ (right-helical electron), and vice versa.

The definitions (24) de facto determines so-called "quantum-mechanical" representation of the Dirac matrices 

\begin{equation}
\label{eq25}
\bar{\gamma}^{\mu}: \, \bar{\gamma}^{\mu}\bar{\gamma}^{\nu}+\bar{\gamma}^{\nu}\bar{\gamma}^{\mu}=2g^{\mu\nu}; \, \bar{\gamma}^{-1}_{0}=\bar{\gamma}_{0}, \, \bar{\gamma}^{-1}_{l}=-\bar{\gamma}_{l},
\end{equation}

\noindent The matrices $\bar{\gamma}^{\mu}$ (25) of this representation are linked to the Dirac matrices $\gamma^{\mu}$ in the standard Pauli-Dirac (PD) representation: 

\begin{equation}
\label{eq26}
\bar{\gamma}^{0}=\gamma^{0}, \, \bar{\gamma}^{1}=\gamma^{1}C, \, \bar{\gamma}^{2}=\gamma^{0}\gamma^{2}C, \, \bar{\gamma}^{3}=\gamma^{3}C, \, \bar{\gamma}^{4}=\gamma^{0}\gamma^{4}C; \quad \bar{\gamma}^{\mu}=v\gamma^{\mu}v, \quad v \equiv\left|
{{\begin{array}{*{20}c}
 \mathrm{I}_{2} \hfill & 0 \hfill \\
 0 \hfill &  C\mathrm{I}_{2} \hfill \\
\end{array} }} \right|=v^{-1},
\end{equation}

\noindent where the standard Dirac matrices $\gamma^{\mu}$ are given by

\begin{equation}
\label{eq27}
\gamma^0 = \left| {{\begin{array}{*{20}c}
 \mathrm{I}_{2} \hfill & 0 \hfill \\
 0 \hfill & -\mathrm{I}_{2} \hfill \\
\end{array} }} \right|, \quad \gamma ^k = \left| {{\begin{array}{*{20}c}
 0 \hfill & {\sigma ^k} \hfill \\
 { - \sigma ^k} \hfill & 0 \hfill \\
\end{array} }} \right|, \quad \mu=0,1,2,3.
\end{equation}

\noindent Note that in the terms of $\bar{\gamma}^{\mu}$ matrices (26) the spin operator (24) have the form $\bar{\overrightarrow{s}}=\frac{i}{4}(\bar{\gamma}^{2}\bar{\gamma}^{3}, \, \bar{\gamma}^{3}\bar{\gamma}^{1}, \, \bar{\gamma}^{1}\bar{\gamma}^{2})$.

The $\bar{\gamma}^{\mu}$ matrices (26) together with the matrix $\bar{\gamma}^{4}\equiv \bar{\gamma}^{0}\bar{\gamma}^{1}\bar{\gamma}^{2}\bar{\gamma}^{3}$, imaginary unit $i\equiv\sqrt{-1}$ and operator $C$ of complex conjugation in $\mathrm{H}^{3,4}$ generate the quantum-mechanical representations of the extended real Clifford-Dirac algebra and proper extended real Clifford-Dirac algebra, which were put into consideration in [9] (see also [10]).

\textbf{On the stationary complete sets of operators}. Let us consider now the outstanding role of the different \textit{complete sets} of operators from the algebra of observables $\mathrm{A}_{\mathrm{S}}$. If one does not appeal to the complete sets of operators, then the solutions of the SF equation (9) are linked directly only with the Sturm-Liouville problem for the energy operator (6). In this case one comes to so-called "degeneration" of solutions. Recall that for an arbitrary complete sets of operators the notion of degeneration is absent in the Sturm-Liouville problem (see, e.g., [5]): only one state vector corresponds to any one point of the common spectrum of a complete set of operators. To wit,	for a comp;ete set of operators there is a one to one correspondence between any point of the common spectrum and an eigenvector.

The \textit{stationary complete sets} (SCS) play the special role among the complete sets of operators. Recall that the SCS is the set of all functionally independent mutually commuting operators, each of which commute with the operator of energy (in our case with the operator (6)). The examples of the SCS in $\mathrm{H}^{3,4}$ are given by $(\widehat{\overrightarrow{p}}, \, s_{z}\equiv s^{3}, \, g)$, $(\overrightarrow{p}, \, \overrightarrow{s}\cdot\overrightarrow{p}, \, g)$, ets. The set $(\overrightarrow{x}, \, s_{z}, \, g)$ is an example of non-stationary complete set. The $\overrightarrow{x}$-realization (5) of the space $\mathrm{H}^{3,4}$ and of quantum-mechanical SF equation (9) are related just to this complete set.

\textbf{The solutions of the Schr$\mathrm{\ddot{o}}$dinger-Foldy equation}. Let us consider the SF equation (9) general solution  related to the SCS $(\widehat{\overrightarrow{p}}, \, \bar{s}_{z}\equiv \bar{s}^{3}, \, g)$, where $\bar{s}^{3}$ is given in (24). The fundamental solutions of the equation (9), which are the eigen solutions of this SCS, are given by the relativistic de Broglie waves:

\begin{equation}
\label{eq28}
\varphi_{\vec{k}\alpha}(t,\overrightarrow{x})=\frac{1}{(2\pi)^{\frac{3}{2}}}e^{-i\omega t + i\vec{k}\vec{x} }\mathrm{D}_{\alpha}, \quad \mathrm{D}_{\alpha}=(\delta^{\beta}_{\alpha}), \quad \alpha = \mathrm{r},\acute{\mathrm{r}}, \quad \mathrm{r}=1,2, \, \acute{\mathrm{r}}=3,4,
\end{equation}

\begin{equation}
\label{eq29}
\mathrm{D}_{\mathrm{r}}\equiv \left| {{\begin{array}{*{20}c}
 \mathrm{d}_{\mathrm{r}}\\
 0\\
\end{array} }} \right|, \, \mathrm{D}_{\acute{\mathrm{r}}}\equiv \left|
{{\begin{array}{*{20}c}
 0\\
 \mathrm{d}_{\acute{\mathrm{r}}}\\
\end{array} }} \right|, \quad \mathrm{d}_{1}=\mathrm{d}_{3}=\left| {{\begin{array}{*{20}c}
 1\\
 0\\
\end{array} }} \right|,\, \mathrm{d}_{2}=\mathrm{d}_{4}=\left|
{{\begin{array}{*{20}c}
 0\\
 1\\
\end{array} }} \right|,
\end{equation}

\noindent where the Cartesian orts $\mathrm{D}_{\alpha}$ are the common eigen vectors for the operators $(\bar{s}_{z}, \, g)$.

Vectors (28) are the generalized solutions of the equation (9). These solutions do not belong to the quantum-mechanical space $\mathrm{H}^{3,4}$, i. e. they are not realized in the nature. Nevertheless, the solutions (28) are the complete orthonormalized orts in the rigged Hilbert space (22). In symbolic form the conditions of orthonormalisation and completeness are given by

\begin{equation}
\label{eq30}
\int d^{3}x\varphi^{\dag}_{\vec{k}\alpha}(t,\overrightarrow{x})\varphi_{\vec{k}^{\prime}\alpha^{\prime}}(t,\overrightarrow{x})=\delta(\overrightarrow{k}-\overrightarrow{k}^{\prime})\delta_{\alpha\alpha^{\prime}},
\end{equation}

\begin{equation}
\label{eq31}
\int d^{3}k\sum_{\alpha=1}^{4}\varphi^{\beta}_{\vec{k}\alpha}(t,\overrightarrow{x})\varphi^{*\beta^{\prime}}_{\vec{k}\alpha}(t,\overrightarrow{x}^{\prime})=\delta(\overrightarrow{x}-\overrightarrow{x}^{\prime})\delta_{\beta\beta^{\prime}}.
\end{equation}

\noindent The functional forms of these conditions are omitted because of their bulkiness.

In the rigged Hilbert space (22) an arbitrary solution of the equation (9) can be decomposed in terms of fundamental solutions (28). Furthermore, for the solutions $f\in\mathrm{S}^{3,4}\subset\mathrm{H}^{3,4}$ the expansion

\begin{equation}
\label{eq32}
f(t,\overrightarrow{x})=\frac{1}{(2\pi)^{\frac{3}{2}}}\int d^{3}xe^{-i\widetilde{k}x}[a^{-}_{\mathrm{r}}(\overrightarrow{k})\mathrm{D}_{\mathrm{r}}+a^{+}_{\acute{\mathrm{r}}}(\overrightarrow{k})\mathrm{D}_{\acute{\mathrm{r}}}^{+}], \quad \widetilde{k}x \equiv \omega t- \overrightarrow{k}\overrightarrow{x}, \quad \omega \equiv \sqrt{\overrightarrow{k}^{2}+m^{2}},
\end{equation}

\noindent is, (i) mathematically well-defined in the framework of the standard differential and integral calculus, (ii) if in the expansion (32) a state $f\in\mathrm{S}^{3,4}\subset\mathrm{H}^{3,4}$, then the amplitudes $(a_{\alpha})=(a^{-}_{\mathrm{r}}, \, a^{+}_{\acute{\mathrm{r}}})$ in (32) belong to the set of the Schwartz test functions over $\mathrm{R}^{3}_{\vec{k}}$. Therefore, they have the unambiguous physical sense of the amplitudes of probability distributions over the eigen values of the SCS $(\widehat{\overrightarrow{p}}, \, \bar{s}_{z}, \, g)$. Moreover, the complete set of quantum-mechanical amplitudes unambiguously determine the corresponding representation of the space $\mathrm{H}^{3,4}$ (in this case it is the $(\overrightarrow{k}, \, \bar{s}_{z}, \, g)$-representation), which vectors have the harmonic time dependence

\begin{equation}
\label{eq33}
\widetilde{f}(t,\overrightarrow{k})=e^{-i\omega t}A(\overrightarrow{k}), \quad A(\overrightarrow{k})\equiv \mathrm{column}(a^{-}_{+},\,a^{-}_{-},\,a^{+}_{-},\,a^{+}_{+}),
\end{equation}

\noindent i. e. are the states with the positive sign of the energy $\widetilde{\omega}$.

The similar assertion is valid for the expansions of the states $f\in\mathrm{H}^{3,4}$ over the basis states, which are the eigen vectors of an arbitrary SCS. Therefore, the corresponding representation of the space $\mathrm{H}^{3,4}$, which is related to such expansions, is often called as the generalized Fourier transformation. 

By the way, the $\overrightarrow{x}$-realization (5) of the states space is associated with the non-stationary complete set of operators $(\overrightarrow{x}, \, s_{z}, \, g)$. Therefore, the amplitudes $f^{\alpha}(t,\overrightarrow{x})=\mathrm{D}^{\dag}_{\alpha}f(t,\overrightarrow{x})=U(t)f(0,\overrightarrow{x})$ of the probability distribution over the eigen values of this complete set depend on time $t$ non-harmonically.

\textbf{On the additional conservation laws}. As it was already mentioned above, the external and internal degrees of freedom  for the free $e^{-}e^{+}$-doublet are independent. Therefore, the operator $\bar{\overrightarrow{s}}$ (24) commutes not only with the operators $\widehat{\overrightarrow{p}}, \overrightarrow{x}$, but also with the orbital part $\widehat{m}_{\mu\nu}$ of the total angular momentum operator. And both operators $\bar{\overrightarrow{s}}$ and $\widehat{m}_{\mu\nu}$ commute with the operator $i\partial_{t}-\sqrt { - \Delta + m^2}$ of the equation (9). Therefore, besides the 10 main (consequences of the 10 Poincar$\mathrm{\acute{e}}$ generators) conservation laws (20), the 12 additional constants of motion exist for the free $e^{-}e^{+}$-doublet. These additional conservation laws are the consequences of the operators of the following observables: 

\begin{equation}
\label{eq34}
\bar{s}_{j}, \, \breve{s}_{ol}=\frac{\bar{s}_{ln}p_{n}}{\widehat{\omega}+m}, \, \widehat{m}_{ln}=x_{l}\widehat{p}_{n}-x_{n}\widehat{p}_{l}, \, \widehat{m}_{0l}=-\widehat{m}_{l0}=t\widehat{p}_{l}-\frac{1}{2}\left\{x_{l},\widehat{\omega}\right\}.
\end{equation}

Thus, the following assertions can be proved. In the space $\mathrm{H}^{\mathrm{A}}=\left\{A\right\}$ of the quantum-mechanical amplitudes the 10 main conservation laws (20) have the form

\begin{equation}
\label{eq35}
(P_{\mu},J_{\mu\nu})=\int d^{3}k A^{\dag}(\overrightarrow{k})(\widetilde{p}_{\mu},\widetilde{j}_{\mu\nu})A(\overrightarrow{k}), \quad A(\overrightarrow{k})\equiv \left|
{{\begin{array}{*{20}c}
 a^{-}_{\mathrm{r}}\\
 a^{+}_{\acute{\mathrm{r}}}\\
\end{array} }} \right|,
\end{equation}

\noindent where the $\mathcal{P}^{\mathrm{A}}$ generators $(\widetilde{p}_{\mu},\widetilde{j}_{\mu\nu})$ of (35) are given by 

\begin{equation}
\label{eq36}
\widetilde{p}_{0}=\omega, \, \widetilde{p}_{l}=k_{l}, \, \widetilde{j}_{ln}=\widetilde{x}_{l}k_{n}-\widetilde{x}_{n}k_{l}+\bar{s}_{ln};\quad (\widetilde{x}_{l}=-i\widetilde{\partial}_{l}, \, \widetilde{\partial}_{l}\equiv \frac{\partial}{\partial k^{l}}),
\end{equation}

\begin{equation}
\label{eq37}
\widetilde{j}_{0l}=-\widetilde{j}_{l0}=-\frac{1}{2}\left\{\widetilde{x}_{l},\omega \right\}-(\breve{\widetilde{S}}_{0l}\equiv\frac{\bar{s}_{ln}k_{n}}{\omega+m}),
\end{equation}

Note that the operators (36), (37) satisfy the Poincar$\mathrm{\acute{e}}$ commutation relations in the manifestly covariant form (3). It is evident that the 12 additional conservation laws (34), generated by the operators (34), are the separate terms in the expressions (35) of total (main) conservation laws.

\textbf{Dynamic and kinematic aspects of the relativistic invariance}. Consider briefly some detalizations of the relativistic invariance of the SF equation (9). Note that for the free $e^{-}e^{+}$-doublet the equation (9) has one and the same explicit form in arbitrary-fixed IFR (its set of solutions is one and the same in every IFR). Therefore, the algebra of observables and the conservation laws (as the functionals of the free $e^{-}e^{+}$-doublet states) have one and the same form too. This assertion explains the dynamical sense of the $\mathcal{P}$ invariance (the invariance with respect to the dynamical symmetry group $\mathcal{P}$).

Another, kinematic, aspect of the $\mathcal{P}$ invariance of the RQCM model has the following physical sense. Note at first that any solution of the SF equation (9) is determined by the concrete given set of the amplitudes $\left\{A\right\}$. It means that if $f$ with the fixed set of amplitudes $\left\{A\right\}$ is the state of the doublet in some arbitrary IFR, then for the observer in the $(a, \, \varpi)$-transformed $\mathrm{IFR}^{\prime}$ this state $f^{\prime}$ is determined by the amplitudes $\left\{A^{\prime}\right\}$. The last ones are received from the given $\left\{A\right\}$ by the unitary  $\mathcal{P}^{\mathrm{A}}$ -transformation 

\begin{equation}
\label{eq38}
\mathcal{P}^{\mathrm{A}}: \, (a,\varpi)\rightarrow \widetilde{U}(a,\varpi)=\exp (-ia^{\mu}\widetilde{p}_{\mu}-\frac{i}{2}\varpi^{\mu\nu}\widetilde{j}_{\mu\nu}),
\end{equation}

\noindent where $(\widetilde{p}_{\mu}, \, \widetilde{j}_{\mu\nu})$ are given in (36), (37).

\textbf{On the principles of the heredity and the correspondence}. The explicit forms (34)-(37) of the main and additional conservation laws demonstrate evidently that the model of RCQM satisfies the principles of the heredity and the correspondence with the non-relativistic classical and quantum theories. The deep analogy between RCQM and these theories for the physical system with the finite number degrees of freedom (where the values of the free dynamical conserved quantities are additive) is also evident.

\textbf{The axiom on the mean value of the operators of observables}. Note that any apparatus can not fulfill the absolutely precise measurement of a value of the physical quantity having continuous spectrum. Therefore, the customary quantum-mechanical axiom about the possibility of "precise" measurement, for example, of the coordinate (or another quantity with the continuous spectrum), which is usually associated with the corresponding "reduction" of the wave-packet, can be revisited. This assertion for the values with the continuous spectrum can be replaced by the axiom that only the mean value of the operator of observable (or the corresponding complete set of observables) is the experimentally observed for $\forall f\in\mathrm{H}^{3,4}$. Such axiom, without any loss of generality of consideration, unambiguously justifies the using of the subspace $\mathrm{S}^{3,4}\subset\mathrm{H}^{3,4}$ as an approximative space of the physically realizable states of the considered object. This axiom as well does not enforce the application of the conception of the ray in $\mathrm{H}^{3,4}$ (the set of the vectors $e^{i\alpha}f$ with an arbitrary-fixed real number $\alpha$) as the state of the object. Therefore, the mapping $(a, \, \varpi)\rightarrow U(a, \, \varpi)$ in the formula (38) and in the formula (18) for the $\mathcal{P}$-representations in $\mathrm{S}^{3,4}\subset\mathrm{H}^{3,4}$ is an unambiguous. Such axiom actually removes the problem of the wave packet "reduction", which discussion started from the well-known von Neumann monograph [11]. Therefore, the subjects of the discussions of all "paradoxes" of quantum mechanics, a lot of attention to which was paid in the past century, are removed also.

The important conclusion about the RCQM is as follows. The consideration of all aspects of this model is given on the basis of using only such conceptions and quantities, which have the direct relation to the experimentally observable physical quantities of this "elementary" physical system.

\textbf{The second quantization}. Finally, we consider briefly the program of the canonical quantization of the RCQM model. Note that the expression for the total energy $P_{0}$ plays the special role in the procedure of so called "second quantization". In the RCQM doublet model, as it is evident from the expression for the $P_{0}$ (35) in the terms of the charge sign-momentum-spin amplitudes 

\begin{equation}
\label{eq39}
P_{0}=\int d^{3}k\omega\left(\left|a^{-}_{\mathrm{r}}(\overrightarrow{k})\right|^{2}+\left|a^{+}_{\acute{\mathrm{r}}}(\overrightarrow{k})\right|^{2}\right)\geq m>0,
\end{equation}

\noindent the energy is positive. The same assertion is valid for the amplitudes related to the arbitrary-fixed SCS of operators. Furthermore, the corresponding to expression (39) operator $\widehat{P}_{0}$ of the energy is positive-valued operator. The operator $\widehat{P}_{0}$ follows from the expression (39) after the anticommutation quantization of the amplitudes

\begin{equation}
\label{eq40}
 \left\{\widehat{a}_{\alpha}(\overrightarrow{k}),\widehat{a}^{\dag}_{\beta}(\overrightarrow{k})\right\}=\delta_{\alpha\beta}\delta\left(\overrightarrow{k}-\overrightarrow{k}^{\prime}\right) 
\end{equation}

\noindent (other operators anticommute) and their substitution $a^{\mp}\rightarrow \widehat{a}^{\mp}$ into the formula (39). Note that the quantized amplitudes determine the Fock space $\mathcal{H}^{\mathrm{F}}$ (over the quantum-mechanical space $\mathrm{H}^{3,4}$). What is more, the operators of dynamical variables $\widehat{P}_{\mu}, \, \widehat{J}_{\mu\nu}$ in $\mathcal{H}^{\mathrm{F}}$, which are expressed according to formulae (35) in the terms of  the operator amplitudes $\widehat{a}_{\alpha}(\overrightarrow{k}), \, \widehat{a}^{\dag}_{\beta}(\overrightarrow{k})$, automatically have the form of "normal products" and satisfy the commutation relations (3) of the $\mathcal{P}$ group in the Fock space $\mathcal{H}^{\mathrm{F}}$. Operators $\widehat{P}_{\mu}, \, \widehat{J}_{\mu\nu}$ determine the corresponding unitary representation in $\mathcal{H}^{\mathrm{F}}$. Other details are not the subject of this paper.

\section{Derivation of the Foldy-Wouthuysen and the standard Dirac equations}

At first we demonstrate that the FW and the standard Dirac equations are the direct and unambiguous consequences of the SF equation (9) and of the RCQM model. Further we consider briefly the physical and mathematical sense of three different models of the Fermionic doublet (RCQM, the FW model, the Dirac model) at the example of comparison of the group-theoretical approaches to these three models.

Thus, below we are looking for the infallible links between the RCQM, the FW model and the Dirac model of the Fermionic doublet. The links between the objects of equations (and between the associated operators of the algebra of observables) of these three models are under consideration. The following mathematical and physical assertions must be taken into account in order to realize such program.

The Poincar$\mathrm{\acute{e}}$ group $\mathcal{P}$ is the group of real parameters $(a, \, \omega)$, i. e. it is the real Lie group. Therefore, as a matter of fact the prime (anti-Hermitian) generators 

\begin{equation}
\label{eq41}
 (p_{\mu}, \, j_{\mu\nu})^{\mathrm{prime}} \equiv (p_{\mu}^{\mathrm{pr}}, \, j_{\mu\nu}^{\mathrm{pr}}) \equiv (-ip_{\mu}, \, -ij_{\mu\nu})
\end{equation}

\noindent are the generators of the group $\mathcal{P}$, and not the operators $(p_{\mu}, \, j_{\mu\nu})$, which were used in all formulae above. In the terms of prime generators (41) the $\mathcal{P}$-representation in the space $\mathrm{H}^{3,4}$ is given by the formula

\begin{equation}
\label{eq42}
 (a,\varpi)\rightarrow U(a,\varpi)=\exp(a^{\mu}p_{\mu}^{\mathrm{pr}}+\frac{1}{2}\varpi^{\mu\nu}j_{\mu\nu}^{\mathrm{pr}}),
\end{equation}

\noindent and the commutation relations of the Lie algebra of the group $\mathcal{P}$ in manifestly covariant form are given by

$$\left[p_{\mu}^{\mathrm{pr}},p_{\nu}^{\mathrm{pr}}\right]=0, \quad \left[p_{\mu}^{\mathrm{pr}},j_{\rho\sigma}^{\mathrm{pr}}\right]=g_{\mu\rho}p_{\sigma}-g_{\mu\sigma}p_{\rho},$$
\begin{equation}
\label{eq43}
\left[j_{\mu\nu}^{\mathrm{pr}},j_{\rho\sigma}^{\mathrm{pr}}\right]=-g_{\mu\rho}j_{\nu\sigma}^{\mathrm{pr}}-g_{\rho\nu}j_{\sigma\mu}^{\mathrm{pr}}-g_{\nu\sigma}j_{\mu\rho}^{\mathrm{pr}}-g_{\sigma\mu}j_{\rho\nu}^{\mathrm{pr}}.
\end{equation}

It was demonstrated in [9], [10] that the prime (anti-Hermitian) generators play the special role in group-theoretical approach for quantum theory and symmetry analysis of the corresponding equations. Just the use of the anti-Hermitian generators of the groups under consideration allowed us [9], [10] to find the additional bosonic properties of the FW and Dirac equations. The mathematical correctness of appealing to the anti-Hermitian generators is considered in details in [12], [13].

Below in all cases we use just the prime (anti-Hermitian) $\mathcal{P}$-generators (and the prime-operators of all energy-momentum an angular momentum quantities are used). Therefore, after this \textit{warning} in the consideration below \textbf{the notation "prime" or "pr" of the corresponding operators is omitted}.  

The link between the SF equation (9) and the FW equation [2] is given by the operator $v$

\begin{equation}
\label{eq44}
 v=\left|
{{\begin{array}{*{20}c}
 \mathrm{I}_{2} \hfill & 0 \hfill \\
 0 \hfill &  C\mathrm{I}_{2} \hfill \\
\end{array} }} \right|; \quad v^{2}=\mathrm{I}_{4}, \quad \mathrm{I}_{2}=\left|
{{\begin{array}{*{20}c}
 1 \hfill & 0 \hfill \\
 0 \hfill & { 1} \hfill \\
\end{array} }} \right|,
\end{equation}

\noindent (we mentioned about the existence of such operator in the formulae (26)), $C$ is the operator of complex conjugation, the operator of involution in the space $\mathrm{H}^{3,4}$. For the non-singular operator $v$ from (44) the equality

\begin{equation}
\label{eq45}
v\left(\partial_{0}+i\widehat{\omega}\right)v=\partial_{0}+i\gamma^{0}\widehat{\omega}
\end{equation}

\noindent is valid. It means that after the transformation $f\rightarrow \phi =vf$ the SF equation (9) becomes the FW equation [2] 

\begin{equation}
\label{eq46}
\left(\partial_{0}+i\gamma^{0}\widehat{\omega}\right)\phi(x)=0; \quad \phi = vf \equiv \left|
{{\begin{array}{*{20}c}
 \varphi_{-} \hfill  \\
 \varphi_{+}^{*} \hfill  \\
\end{array} }} \right|\in \mathrm{H}^{3,4}, \quad x\equiv (t,\overrightarrow{x}).
\end{equation}

\textbf{The Foldy-Wouthuysen model of the Fermionic doublet}. The quantum-mechanical sense of the object $\phi$ is following. The equation (46) is the system of two 2-component equations

\begin{equation}
\label{eq47}
(\partial_{0}+i\widehat{\omega})\varphi_{-}(x)=0, \quad (\partial_{0}-i\widehat{\omega})\varphi_{+}^{*}(x)=0.
\end{equation}

\noindent First equation is the equation for the wave function $\varphi_{-}$ of the electron and second -- for the function $\varphi_{+}^{*}$ being the complex conjugate to the wave function $\varphi_{+}$ of the positron.

Under the transformations $f\rightarrow \phi =vf$ the $\mathcal{P}^{\mathrm{f}}$-generators (14), (15) \textit{(taken with the spin term (24) and in the prime form)} become the prime $\mathcal{P}^{\mathrm{\phi}}$-generators ($\mathcal{P}$-symmetries of the FW equation (46))

\begin{equation}
\label{eq48}
\widehat{p}_{0}=-i\gamma_{0}\widehat{\omega}, \, \widehat{p}_{\ell}=\partial_{\ell}, \, \widehat{j}_{\ell n}=x_{\ell}\partial_{n}-x_{n}\partial_{\ell}+\widehat{s}_{\ell n}\equiv \widehat{m}_{\ell n}+\widehat{s}_{\ell n},
\end{equation}

\begin{equation}
\label{eq49}
\widehat{j}_{0\ell}=t\partial_{\ell}+\frac{i}{2}\gamma_{0}\left\{x_{l},\widehat{\omega}\right\}+\gamma_{0}\frac{\widehat{s}_{\ell n}\widehat{p}_{n}}{\widehat{\omega}+m}; \quad \widehat{s}_{\ell n}=\widehat{s}^{\ell n}\equiv\frac{1}{4}\left[\gamma^{\ell},\gamma^{n}\right],
\end{equation}

\noindent where $\gamma^{\mu}$ are the standard Dirac matrices (27) (in the Pauli -- Dirac representation).

Therefore, $\mathcal{P}^{\mathrm{\phi}}$-generators (48), (49) of the $\mathcal{P}^{\mathrm{\phi}}$-representation in $\mathrm{H}^{3,4}$ (as well as the operators $q^{\mathrm{\phi}}=vq^{\mathrm{f}}v$) of the algebra of all observable physical quantities in the FW model of the Fermionic doublet) are the functions generated by the 9 operators

\begin{equation}
\label{eq50}
\overrightarrow{x}=(x^{j})\in \mathrm{R}^{3}, \, \widehat{\overrightarrow{p}}=-\nabla, \, \widehat{\overrightarrow{s}}\equiv (\widehat{s}_{23},\widehat{s}_{31},\widehat{s}_{12}),
\end{equation}

\noindent where $\widehat{s}_{\ell n}$ are given in (49). The physical sense of these operators (as well as of the functions $q^{\mathrm{\phi}}$ from them) follows from the physical sense of the corresponding quantum-mechanical operators (21) (and $q^{\mathrm{f}}$), which are verified by the principles of the heredity and correspondence with non-relativistic quantum and classical theories.

The solution (32) of the SF equation (9), which is associated with the SCS (momentum -- sign of the charge -- spin projection $\bar{s}_{\mathrm{z}}$) transforms into the following solution of the FW equation (46)

\begin{equation}
\label{eq51}
\phi(x)=\frac{1}{(2\pi)^{\frac{3}{2}}}\int d^{3}x\left(e^{-i\widetilde{k}x}a^{-}_{\mathrm{r}}(\overrightarrow{k})\mathrm{D}_{\mathrm{r}}+e^{i\widetilde{k}x}a^{*+}_{\acute{\mathrm{r}}}(\overrightarrow{k})\mathrm{D}_{\acute{\mathrm{r}}}^{+}\right), 
\end{equation}

\noindent where $a^{-}_{\mathrm{r}}(\overrightarrow{k})$ are the same as in (32) and $a^{*+}_{\acute{\mathrm{r}}}(\overrightarrow{k})$ are the complex conjugated functions to the quantum-mechanical amplitudes $a^{+}_{\acute{\mathrm{r}}}(\overrightarrow{k})$ of the standard quantum-mechanical sense, which was considered in the previous section.

Further, similarly to the RCQM in the FW model the additional conservation laws also exist together with the main 10 Poincar$\mathrm{\acute{e}}$ conservation quantities 

\begin{equation}
\label{eq52}
(\widehat{p}_{\mu},\widehat{j}_{\mu\nu})^{\mathrm{\phi}} \rightarrow (P_{\mu},J_{\mu\nu})^{\mathrm{\phi}}=\int d^{3}x \phi^{\dag}(x)i(\widehat{p}_{\mu},\widehat{j}_{\mu\nu})^{\mathrm{\phi}}\phi(x).
\end{equation}

\noindent The 12 additional conservation laws, which were considered in the section 2, also exist and con be very easy calculated here. Naturally, due to non-unitarity of the operator $v$ from (44) the explicit form of the conservation laws (52) does not coincide with the quantum-mechanical quantities (35). It is evident from the expression (52) in the terms of quantum-mechanical amplitudes

\begin{equation}
\label{eq53}
(P_{\mu},J_{\mu\nu})^{\mathrm{\phi}}=\int d^{3}k A^{\mathrm{\phi}\dag}(\overrightarrow{k})(\widetilde{p}_{\mu},\widetilde{j}_{\mu\nu})^{\mathrm{\phi}}A^{\mathrm{\phi}}(\overrightarrow{k}); \quad A(\overrightarrow{k})\equiv \left|
\begin{array}{cccc}
 a^{-}_{+} \\
 a^{-}_{-} \\
a^{*+}_{+} \\
a^{*+}_{-} \\
\end{array} \right|,
\end{equation}

\noindent where $(\widetilde{p}_{\mu},\widetilde{j}_{\mu\nu})^{\mathrm{\phi}}$ are given by

\begin{equation}
\label{eq54}
\widetilde{p}_{0}=\gamma^{0}\omega, \, \widetilde{p}_{\ell}=\gamma^{0}k_{\ell}, \, \widetilde{j}_{\ell n}=\widetilde{x}_{\ell}k_{n}-\widetilde{x}_{n}k_{\ell}+\widehat{s}_{\ell n};\quad (\widetilde{x}_{\ell}=-i\widetilde{\partial}_{\ell}, \, \widetilde{\partial}_{\ell}\equiv \frac{\partial}{\partial k^{\ell}}),
\end{equation}

\begin{equation}
\label{eq55}
\widetilde{j}_{0\ell}=-\widetilde{j}_{\ell 0}=-\frac{1}{2}\left\{\widetilde{x}_{\ell},\omega \right\}+\gamma^{0}(\breve{\widetilde{S}}_{0\ell}\equiv\frac{\widehat{s}_{\ell n}k_{n}}{\omega+m}); \quad \omega \equiv \sqrt{\overrightarrow{k}^{2}+m^{2}}.
\end{equation}

For example, the total energy of the field $\phi$, instead of the expression (39) in RCQM, has the form

\begin{equation}
\label{eq56}
P_{0}=\int d^{3}k \omega \left(a^{*-}_{\mathrm{r}}(\overrightarrow{k})a^{-}_{\mathrm{r}}(\overrightarrow{k})-a^{+}_{\acute{\mathrm{r}}}(\overrightarrow{k})a^{*+}_{\acute{\mathrm{r}}}(\overrightarrow{k})\right),
\end{equation}

\noindent which is not positive defined. In this sense, the FW model (which quantum-mechanical content is unambiguous) in fact is not the quantum-mechanical model for the $e^{-}e^{+}$-doublet. Therefore, in the procedure of "canonical quantization" of the field $\phi$ on the basis of anticommutation relations (40) the \textit{additional axiom} is applied in this model for the quantized field $\widehat{\phi}$. According to this axiom the definition of the operators of the physical quantities in the Fock space $\mathcal{H}^{\mathrm{F}}$ is extended by taking them only in the form of "normal products" with respect to the operator amplitudes $\widehat{a}^{-}_{\mathrm{r}}(\overrightarrow{k}), \, \widehat{a}^{*-}_{\mathrm{r}}(\overrightarrow{k}), \, \widehat{a}^{+}_{\acute{\mathrm{r}}}(\overrightarrow{k}), \, \widehat{a}^{*+}_{\acute{\mathrm{r}}}(\overrightarrow{k})$. It is easy to verify that the operators of 10 main conserved quantities of the "quantized field" $\widehat{\phi}$ in the form of normal products coincide with the corresponding expressions in the "second quantized" RCQM model of the Fermionic doublet

\begin{equation}
\label{eq57}
:(\widehat{P}_{\mu},\widehat{J}_{\mu\nu})^{\mathrm{\phi}} :=\int d^{3}k \widehat{A}^{\dag}(\overrightarrow{k})(\widetilde{p}_{\mu},\widetilde{j}_{\mu\nu})\widehat{A}(\overrightarrow{k}); \quad \widehat{A}(\overrightarrow{k})\equiv \left|
\begin{array}{cccc}
 \widehat{a}^{-}_{+} \\
 \widehat{a}^{-}_{-} \\
\widehat{a}^{+}_{-} \\
\widehat{a}^{+}_{+} \\
\end{array} \right|,
\end{equation}

\noindent where $(\widetilde{p}_{\mu},\widetilde{j}_{\mu\nu})$ are given in (36), (37).

\textbf{The Dirac model of the Fermionic doublet}. Taking into account the consideration above and using the well-known from [2] transition operators $V^{\pm}$

\begin{equation}
\label{eq58}
 \phi\rightarrow\psi=V^{+}\phi, \, \psi\rightarrow\phi=V^{-}\psi, \quad V^{\pm}\equiv\frac{\pm i\gamma^{l}\partial_{l}+\widehat{\omega}+m}{\sqrt{2\widehat{\omega}(\widehat{\omega}+m)}},
\end{equation}

\noindent we find the resulting operator

\begin{equation}
\label{eq59}
 W=V^{+}v, \, W^{-1}=vV^{-}: \quad f\rightarrow\psi=Wf, \, \psi\rightarrow f= W^{-1}\psi \quad WW^{-1}=W^{-1}W=1,
\end{equation}

\noindent which transforms (one-to-one) all quantities of RCQM model into the corresponding quantities of the Dirac model and vice versa. For example,

\begin{equation}
\label{eq60}
 W\left(\partial_{0}+i\widehat{\omega}\right)W^{-1}=\partial_{0}+iH_{\mathrm{D}}; \quad H_{\mathrm{D}}\equiv \overrightarrow{\alpha}\cdot \overrightarrow{p}+\beta m.
\end{equation}

\noindent It means that quantum-mechanical SF equation (9) transforms into the Dirac equation in the Schr$\mathrm{\ddot{o}}$dinger form 

\begin{equation}
\label{eq61}
 \left(\partial_{0}+i\gamma^{0}\widehat{\omega}\right)\phi(t,\overrightarrow{x})=0 \rightarrow \left(\partial_{0}+i(\overrightarrow{\alpha}\cdot \overrightarrow{p}+\beta m)\right)\psi(t,\overrightarrow{x})=0.
\end{equation}

Furthermore, 

\begin{equation}
\label{eq62}
 W(\widehat{p}_{\mu},\widehat{j}_{\mu\nu})W^{-1}=(\breve{p}_{\mu},\breve{j}_{\mu\nu})^{\mathrm{Dirac} } ,
\end{equation}

\noindent where

\begin{equation}
\label{eq63}
\breve{p}_{0}=-iH_{\mathrm{D}}, \, \breve{p}_{\ell}=\partial_{\ell}, \, \breve{j}_{\ell n}=x_{\ell}\partial_{n}-x_{n}\partial_{\ell}+\widehat{s}_{\ell n}\equiv \widehat{m}_{\ell n}+\widehat{s}_{\ell n},
\end{equation}

\begin{equation}
\label{eq64}
\breve{j}_{0\ell}=t\partial_{\ell}+\frac{i}{2}\left\{x_{l},H_{\mathrm{D}}\right\}=t\partial_{\ell}-x_{\ell}\breve{p}_{0}+\widehat{s}_{0 \ell}; \quad \widehat{s}_{\mu\nu}\equiv\frac{1}{4}\left[\gamma_{\mu},\gamma_{\nu}\right].
\end{equation}

Note that the generators of the \textit{local} $\mathcal{P}^{\mathrm{\psi}}$-representation (the standard $\mathcal{P}$-algebra of invariance of the Dirac equation (61)) have the form

\begin{equation}
\label{eq65}
p_{\mu}= \partial _{\mu}, \quad j_{\mu\nu}= x_{\mu}\partial _{\nu}-x_{\nu}\partial_{\mu}+\widehat{s}_{\mu\nu}.
\end{equation}

\noindent Moreover, in the manifold ${\psi}\subset\mathrm{H}^{3,4}$ of the solutions of the Dirac equation (61) operators (65) coincide with the $\mathcal{P}^{\mathrm{I}}$-generators (63), (64). Therefore, the $\mathcal{P}^{\mathrm{\psi}}$-representation determined by the generators (63), (64) is called induced (I). As a consequence of this fact, for example, the following equalities are valid for the 10 main dynamical variables of the field $\psi$ 

\begin{equation}
\label{eq66}
(p_{\mu}, j_{\mu\nu})\rightarrow(P_{\mu},J_{\mu\nu})^{\mathrm{\psi}}\equiv \int d^{3}x \overline{\psi}i(p_{\mu}, j_{\mu\nu})\psi = \int d^{3}x \overline{\psi}i(\breve{p}_{\mu}, \breve{j}_{\mu\nu})\psi = (P_{\mu},J_{\mu\nu})^{\mathrm{\phi}},
\end{equation}

\noindent where the last conservation laws are given by (52) and in the terms of amplitudes $a^{-}_{\mathrm{r}}(\overrightarrow{k}), \, a^{*-}_{\mathrm{r}}(\overrightarrow{k}), \, a^{+}_{\acute{\mathrm{r}}}(\overrightarrow{k}), \, a^{*+}_{\acute{\mathrm{r}}}(\overrightarrow{k})$ -- by the formula (53). It means that the procedure of canonical quantization of the field $\psi$ is reduced to the corresponding procedure of the field $\phi$ quantization.

We pay attention that $\mathcal{P}$-operators (65) are the functions of the 14 independent "generated" operators $x_{\mu}, \ \partial_{\mu}, \, \widehat{s}_{\mu\nu}$. Further, $\mathcal{P}$-generators (63), (64) are the functions of 12 independent operators $x_{\ell}, \ \partial_{\ell}, \, \widehat{s}_{\mu\nu}$. Nevertheless, only the operator $\overrightarrow{p}=-\nabla$ has the physical sense of the quantum-mechanical Fermi doublet momentum operator among the above mentioned 14 independent operators. As it was proved in [2] the operators $\overrightarrow{x}=(x^{\ell})$ and $\widehat{s}_{\mu\nu}$, which are essentially used in the constructions (63)--(65), do not have the physical sense of the quantum-mechanical operators of the Fermionic doublet coordinate and SU(2)-spin. This fact evidently demonstrates the validity of the assertion that the standard Dirac local model is not the \textit{quantum-mechanical model} of the fermionic doublet at all.

\section{Conclusions}

The model of relativistic canonical quantum mechanics on the level of axiomatic approaches to the quantum field theory is considered. The main intuitive physical principles, reinterpreted on the level of modern physical methodology, mathematically correctly are mapped into the basic assertions (axioms) of the model. The Einstein's principle of relativity is mapped as a requirements of special relativity. The principles of heredity and correspondence of the model with respect to the non-relativistic classical and quantum mechanics are supplemented by the clarifications of external and internal degrees of freedom carriers. The principle of relativity of the model with respect to the means of cognition is realized by the applications of the rigged Hilbert space. The Schwartz test function space $\mathrm{S}^{3,4}$ is shown to be the sufficient to satisfy the requirements of the principle of relativity of the model with respect to the means of cognition. And the fulfilling of calculations in $\mathrm{S}^{3,4}$ does not lead to the loss of generality of the consideration.

It is shown that the algebra of experimentally observed quantities, associated with the Poincar$\mathrm{\acute{e}}$-invariance of the model, is determined by the nine functionally independent operators $\overrightarrow{x}, \overrightarrow{p}, \overrightarrow{s}$, which in the relativistic canonical quantum mechanics model of the doublet have the unambiguous physical sense. It is demonstrated that the application of the stationary complete sets of operators of the experimentally measured physical quantities guarantees the visualization and the completeness of the consideration. 

Derivation of the Foldy-Wouthuysen and the Dirac equations from the Schr$\mathrm{\ddot{o}}$dinger-Foldy equation of relativistic canonical quantum mechanics is presented and briefly discussed. We prove that the Dirac equation is the consequence of more elementary model of the same physical reality. The relativistic canonical quantum mechanics is suggested to be such fundamental model of the physical reality. Moreover, it is suggested to be the most fundamental model of the Fermi spin $s=\frac{1}{2}$-doublet.

\textbf{An important assertion is that} \textit{an arbitrary physical and mathematical information, which contains in the model of relativistic canonical quantum mechanics, is translated directly and unambiguously into the information of the same physical content in the field model of the Dirac equation.}

Hence, the Dirac equation is the unambiguous consequence of the relativistic canonical quantum mechanics of the Fermi spin $s=\frac{1}{2}$-doublet (e. g. $e^{-}e^{+}$-doublet). Nevertheless, the model of relativistic canonical quantum mechanics of the Fermi-doublet has evident independent application. As a brief example we consider the application of our model to the idea of a possible negative mass of the antiparticle. The negative mass of antiparticle was considered, for example, in [14]-[17], and also is the subject of recent experimental investigations. 

It is interesting to note that the paper [15] contains some theoretical reasons why  the mass of the antiparticle (e. g. positron) must be taken negative, $m_{+}<0$ 	
in contradistinction to the mass of the particle (e. g. electron), which is positive, $m_{-}>0$. We pay attention that the model of the relativistic canonical quantum mechanics of the Fermi-doublet does not need the application of the positron negative mass concept. It is natural due to the following reasons. It is only the energy, which depends from the mass. And the energy together with the momentum is associated with the external degrees of freedom, which are common and the same for the particle and antiparticle (for the electron and positron). The difference between $e_{-}$ and $e_{+}$ contains only in internal degrees of freedom such as the spin $\overrightarrow{p}$ and sign of the charge $g=-\gamma^{0}$. Thus, if in the relativistic canonical quantum mechanics the mass of the particle is taken positive then the mass of the antiparticle must be taken positive too.

On the other hand the comprehensive analysis [15] of the Dirac equation for the doublet had led the authors of the article [15] to the concept of the negative mass of the antiparticle. Therefore, our consideration in the last paragraph gives the additional arguments that the Dirac (or associated with it Foldy-Wouthuysen) model is not the quantum-mechanical ones. Furthermore, in the problem of the relativistic hydrogen atom the use of negative-frequency part $\psi^{-}(x)=e^{-i\omega t}\psi(\overrightarrow{x})$ of the spinor $\psi(x)$ in the "role of the quantum-mechanical object" is not a valid. In this case neither $\left|\psi(\overrightarrow{x})\right|^{2}$, nor $\overline{\psi}(\overrightarrow{x})\psi(\overrightarrow{x})$ is the probability distribution density with respect to the eigen values of the Fermi-doublet coordinate operator. It is due to the fact [2] that in the Dirac model the $\overrightarrow{x}$ is not the Fermi-doublet coordinate operator.

The aplication of the relativistic canonical quantum mechanics can be useful for the analysis of the experimental situation found in [18]. Such analysis is interesting due to the fact that (as it is demonstrated here in the sectiones 2 and 3) the relativistic canonical quantum mechanics is the most fundamental model of the Fermi-doublet. 

Another interesting application of the relativistic canonical quantum mechanics is inspired by the article [19], where the quantum electrodynamics is reformulated in the Foldy-Wouthuysen representation. The author of [19] essentially used the result of the [15] about the negative mass of the antiparticle. Starting from the relativistic canonical quantum mechanics we are able not to appeal to the conception of the antiparticle negative mass.

\vskip 1.cm

\end{document}